\documentclass[aps,pra,amssymb, amsfonts,amsmath,showpacs]{revtex4}
\begin{document}
\title{Stress Tensor for Quantized Random Field and Wave Function Collapse}
\author{Philip Pearle}
\email{ppearle@hamilton.edu}
\affiliation{Department of Physics, Hamilton College, Clinton, NY  13323}
\date{\today}
\begin{abstract}
{The continuous spontaneous localization (CSL) theory of dynamical wave function collapse is an experimentally testable alternative to non-relativistic quantum mechanics.  In it, collapse occurs because particles interact with a classical random field.  However, particles gain energy from this field, i.e., particle energy is not conserved.  Recently, it has been shown how to construct a theory dubbed  ``completely quantized collapse" (CQC) which is predictively equivalent to CSL.  In CQC, a quantized random field is introduced, and CSL's classical random field becomes its eigenvalue.  In CQC, energy is conserved, which allows one to understand that energy is conserved in CSL, as the particle's energy gain is compensated by the random field's energy loss.  Since the random field has energy, it  should have gravitational consequences.   For that, one needs to know the random field's energy density.  In this paper, it is shown how to construct a symmetric, conserved, energy-momentum-stress-density tensor associated with the quantized random field, even though this field obeys no dynamical equation and has no Lagrangian.  Then, three examples are given involving the random field's energy density. One considers interacting particles, the second treats a ``cosmological" particle creation model,  the third involves the  gravity of the random field.}

\end{abstract}\label{Section 0}
\maketitle
\section{Introduction}
The CSL (Continuous Spontaneous Localization) theory\cite{CSL}  describes wave function collapse as a physical, dynamical,  process.   A motivation for thus altering standard quantum theory is that CSL has a well-defined interpretation.  By comparison, no one has been able to graft a well-defined interpretation onto the formalism of standard quantum theory.  A crucial test for such an interpretation is that it give rules for specifying the realizable (i.e., the possible) states of nature, and their probabilities of realization. 
There are no such rules for standard quantum theory: people decide upon the possible states in ad hoc fashion, utilizing educated judgments about specific physical situations.  
	
	In CSL, the Schr\"odinger equation is modified by adding a term containing a classical white-noise field  $w({\bf x},t)$.   {\it Every} state evolving under a possible field is a possible state.  However, the theory also provides a rule for applying a probability measure to each random field.  The result is that only a  subset of  random fields evolve {\it probable} states.  The dynamics is such that a superposition of states, which differ sufficiently in their mass density distributions, rapidly evolves to one such state.   That is, the theory describes the macroscopic  world we see around us.  In many cases, CSL provides the same experimental predictions obtained from standard quantum theory, but there are experiments which can distinguish between them\cite{experiments}.  As emphasized by Bohm\cite{Bohm}, this possibility of suggesting and stimulating experiments, which might not otherwise be envisaged or undertaken, provides a strong reason for exploring the consequences of an alternative to quantum theory, were there no other reason.  
	
	One experimentally testable consequence of CSL is that particles gain energy, e.g. electrons in atoms and nucleons in nucleii should  ``spontaneously"  be excited.  Experiments have show that the white-noise-particle coupling has to be mass-proportional, or otherwise a discrepancy would have been observed.  This is suggestive of a connection between collapse and gravity, which a number of authors have proposed for other reasons\cite{gravity}. 
	
	Energy non-conservation in a theory is not considered to be a good thing.  However, recently, 
a theory called CQC (Completely Quantized Collapse) has been investigated which, it has been shown, is completely equivalent to CSL\cite{CQC}.  CQC has in it a quantized random field $W({\bf x},t)$, which commutes with itself everywhere in space-time, and its eigenvalues are the CSL $w({\bf x},t)$.  Since it is a Hamiltonian theory, energy is conserved: the increase in particle energy is precisely compensated by the decrease in the random field's energy, and this is then true of CSL as well. 

But, with these new ideas, there is the potential for new physics.  A new source of energy ought to have new gravitational effects.  However, to calculate those effects requires knowing the energy density.  The purpose of this paper is to supply an expression for the energy density.  

	In Section II we define $W({\bf x},t)$.  Section III defines its energy-momentum.  Section IV (and Appendix A) presents a symmetric conserved energy-momentum-stress density tensor whose spatial integral equals its  energy-momentum. 
	
	Section V gives a brief introduction to CSL, with pertinent equations. Section VI contains a discussion of CQC, and explains how conclusions about CSL can be drawn from CQC, in particular, how to express the stress tensor in the language of CSL.  
	
	Section VII and VIII discuss collapse dynamics for non-relativistic, interacting particles. Expressions for the ensemble-average particle and random field energy-densities are given.  Section IX gives the comparable expressions for a model of`cosmogenesis, with a Hamiltonian which creates particles  out of the vacuum.  
	
	Section X presents an example where the random field energy density interacts gravitationally with the particle mass-density. Collapse dynamics is considered for a superposition of static particle mass-density distributions.  The result is that collapse occurs as usual, except that the state vector at time $t$ describes the state of the particles on the proper time hypersurface  $t'=t[1+\phi ({\bf x}')]\approx t\sqrt{-g_{00}({\bf x}')}$ ($t'$,  ${\bf x}'$ are the space-time coordinates, $\phi ({\bf x}')$ is the gravitational potential, and $g_{00}({\bf x}')= -1-2\phi ({\bf x}')$ is the 00th component of the metric tensor in the limit of a weak gravitational field).   

\section{Quantized White Noise Field}\label{Section I}

We define a quantized white noise field 
\begin{equation}\label{1}
W(x)\equiv\frac{\sqrt{\lambda}}{(2\pi)^{2}}\int dk [e^{ik\cdot x} b(k)+
e^{-ik\cdot x}b^{\dagger}(k)], 
\end{equation}
where $\lambda$ is a constant, $x^{\mu}=({\bf x}, t)$, $k^{\mu}=({\bf k}, k^{0})$, $k\cdot x={\bf k\cdot x}-k^{0}t$, and 
\begin{equation}\label{2}
[b(k),b^{\dagger}(k')]=\delta(k-k'), \quad [b(k),b(k')]=0, \quad [b^{\dagger}(k),b^{\dagger}(k')]=0. 
\end{equation}
We also define a conjugate field 
\begin{equation}\label{3}
\Pi(x)\equiv\frac{i}{2\sqrt{\lambda}(2\pi)^{2}}\int dk [-e^{ik\cdot x} b(k)+
e^{-ik\cdot x}b^{\dagger}(k)].  
\end{equation}
It easily follows that 
\begin{equation}\label{4}
 [W(x),\Pi(x')]= i\delta(x-x'), \quad [W(x),W(x')]=0, \quad [\Pi(x),\Pi(x')]=0. 
\end{equation}

	For fixed $x$, $W(x)$ has an eigenvector with eigenvalue w(x) which can take on any value between $-\infty$ and $\infty$.  Because $W(x)$ has the unusual property for a quantized field that it commutes with itself everywhere in spacetime, a joint eigenvector of  $W(x)$ for every $x$ can be constructed,  	
\begin{equation}\label{5}
W(x)|w\rangle=w(x)|w\rangle,
\end{equation}
Since $w(x)$ can take on any value at any $x$, it is a sample classical white noise field.  

	The vacuum state $|0\rangle$ satisfies $\int dkf(k)b(k)|0\rangle=0$, where $f(k)$ is an arbitrary function.  Therefore,  from (\ref{1}) and (\ref{3}), $[W(x)+2i\lambda\Pi (x)]|0\rangle=0$, or 
\begin{equation}\label{6}
\langle w|W(x)+2i\lambda\Pi(x)|0\rangle=[w(x)+2\lambda\delta/\delta w(x)]\langle w|0\rangle=0,
\end{equation}
( $\delta/\delta w(x)$ is the functional derivative).  The solution of (\ref{6}) is
\begin{equation}\label{7}
\langle w|0\rangle=e^{-(4\lambda)^{-1}\int_{-\infty}^{\infty}dxw^{2}(x)}.  \end{equation}
The limits on integrals over ${\bf x}$ in this paper are always $(-\infty, \infty)$, so an explicit limit, as in Eq. (\ref{7}), always refers to $t$.  The states $\langle w |$ are summed over utilizing     $Dw\sim\prod_{x}dw(x)$, where the constant  is chosen so that $\int_{-\infty}^{\infty}Dw|w\rangle\langle w|=1$ (a  rigorous definition can be given by discretizing space-time\cite{CQC}), e.g.,  
$\int Dw\langle 0|w\rangle\langle w|0\rangle=\langle 0|0\rangle=1$.

\section{$W$-Field Energy-Momentum}
	
	The  four-momentum, which generates space-time translations of $W(x)$, $\Pi (x)$, is 
\begin{equation}\label{8}
P_{w}^{\nu}\equiv\int dkk^{\nu} b^{\dagger}(k)b(k).
\end{equation}
By solving Eqs.(\ref{1}), (\ref{3}) for $b(k)$, we find
\begin{equation}\label{9}
b(k)=\frac{1}{(2\pi)^{2}}\int dx\Big[\frac{1}{2\sqrt{\lambda}}W(x)+i\sqrt{\lambda}\Pi(x)\Big]e^{-ik\cdot x}.  
\end{equation}
Substituting (\ref{9}) into (\ref{8}), using $k^{\nu}\exp ik\cdot x=-i(\partial/\partial x_{\nu})\exp ik\cdot x$ in the expression for $b^{\dagger}$, and integrating over $k$, there results:	
\begin{subequations}\label{10} 
\begin{eqnarray}
P_{w}^{\nu}&\equiv&-i\int dx \frac{\partial}{\partial x_{\nu}}\Big[\frac{1}{8\lambda}W^{2}(x)+\frac{\lambda}{2} \Pi^{2}(x)\Big]+
\int dx\frac{1}{2}\big[W(x)\frac{\partial}{\partial x_{\nu}}\Pi(x)-\Pi(x)\frac{\partial}{\partial x_{\nu}}W(x)\Big]\\
&=&\int dx\frac{1}{2}\big[W(x)\frac{\partial}{\partial x_{\nu}}\Pi(x)-\Pi(x)\frac{\partial}{\partial x_{\nu}}W(x)\Big]\\
&=&\int dx W(x)\frac{\partial}{\partial x_{\nu}}\Pi(x)=-\int dx \Pi(x)\frac{\partial}{\partial x_{\nu}}W(x)=
-\int dx\Big[\frac{\partial} {\partial x_{\nu}}W(x)\Big]\Pi(x)\\
&=&\int dx\{\dot W(x), -{\bf \nabla} W(x)\}\Pi(x).
\end{eqnarray}
\end{subequations} 

In going from (10a) to (10b), we have assumed no contribution from the boundary terms.  This may be achieved e.g., by defining $W(x)$, $\Pi(x)$  to vanish outside a suitably large space-time hypervolume.   
 
 In (10c), we also recognize that the 
 operators can be commuted freely within the integral, since $\partial\delta(x-x)/\partial x_{\nu}=0$. 
 
  Eq.(10d) 
 displays $\{P_{w}^{0},{\bf P}_{w}\}$.   
 
  \section{$W$-Field Energy-Momentum-Stress Density Tensor}
 We want to construct a stress tensor $T_{w}^{\mu\nu}(x)$ for the $W$-field, consistent with the constraints that it is symmetric, satisfies  $\partial_{\nu}T_{w}^{\mu\nu}(x)=0$ and $P_{w}^{\nu}=\int d{\bf x}T_{w}^{0\nu}(x)$.  It is well known that stress tensors are not necessarily unique. For example, one can add a term to a field theory Lagrangian,  of the form of a space or time derivative of a function of fields, which does not change the equations of motion but which does change the canonically constructed stress tensor.  Therefore, we shall be content to display a stress tensor which satisfies these constraints.  
 
 Appendix A contains a "derivation" of this stress tensor expression. Indeed,  there appears an added term which is dropped without violating the above constraints.  Not only does this dropped term make no contribution to the energy-momentum, but it creates and annihilates ``W-particles," whereas the  $T_{w}^{\mu\nu}(x)$ which remains and is displayed here does not. 

   Without further ado:
  \begin{subequations}\label{11} 
 \begin{eqnarray}
T_{w}^{\mu\nu}(x)&\equiv&\frac{1}{2(2\pi)^{7}}\int dx_{1}dx_{2} \int dk_{1} dk_{2}
\delta (k_{1}^{2}-k_{2}^{2})\sum_{s=-1}^{1}s\Theta (sk_{1}^{0})\Theta (sk_{2}^{0})\sin [k_{1}\cdot(x-x_{1})-k_{2}\cdot(x-x_{2})][\nonumber\\
&&\cdot[k_{1}^{\mu}k_{2}^{\nu}+k_{1}^{\nu}k_{2}^{\mu}+(1/2)\eta^{\mu\nu}(k_{1}-k_{2})^{2}]
W(x_{1})\Pi(x_{2})-\Pi(x_{1})W(x_{2})]\\
&=&\frac{1}{2(2\pi)^{7}}\int dx_{1}dx_{2} \int dk_{1} dk_{2}
\delta (k_{1}^{2}-k_{2}^{2})\sum_{s=-1}^{1}s\Theta (sk_{1}^{0})\Theta (sk_{2}^{0})\sin [k_{1}\cdot(x-x_{1})-k_{2}\cdot(x-x_{2})]\nonumber\\
&&\cdot\Bigg[\frac{\partial}{\partial x_{1\mu}}\frac{\partial}{\partial x_{2\nu}}+\frac{\partial}{\partial x_{2\mu}}\frac{\partial}{\partial x_{1\nu}}-(1/2)\eta^{\mu\nu}
(\frac{\partial}{\partial x_{1}}+\frac{\partial}{\partial x_{2}})^{2}\Bigg]
[W(x_{1})\Pi(x_{2})-\Pi(x_{1})W(x_{2})],
\end{eqnarray}
\end{subequations}
 In (\ref{11}), $\Theta$ is the step function and the metric tensor's non-vanishing diagonal 
 elements are $\eta^{\mu\mu}=(-1,1,1,1)$.  
 
 We note that $T_{w}^{\mu\nu}(x)=T_{w}^{\nu\mu}(x)$.  Now, 
 consider $\partial_{\nu}T_{w}^{\mu\nu}(x)$.  The derivative, acting on the 
 sine, results in a factor $ [k_{1\nu}-k_{2\nu}]$.  Then, factors in the integrand are
 \[ \delta (k_{1}^{2}-k_{2}^{2})[k_{1\nu}-k_{2\nu}][k_{1}^{\mu}k_{2}^{\nu}+k_{1}^{\nu}k_{2}^{\mu}+(1/2)\eta^{\mu\nu}(k_{1}-k_{2})^{2}]=\delta (k_{1}^{2}-k_{2}^{2})(1/2)[k_{1}^{\mu}+k_{2}^{\mu}][k_{1}^{2}-k_{2}^{2}]=0
\]
so $\partial_{\nu}T_{w}^{\mu\nu}(x)=0$. It is remarkable that this 
equation describing the local flow of four-momentum is satisfied, 
given the very non-local nature of $T_{w}^{\mu\nu}(x)$ displayed in  Eqs.(\ref{13}-\ref{15}) below.  
 
Finally, consider the spatial integral of (11a): 
  \begin{subequations}\label{12} 
 \begin{eqnarray}
 \int d{\bf x}T_{w}^{0\nu}(x)&=&\frac{1}{2(2\pi)^{4}}\int dx_{1}dx_{2} \int dk_{1} dk_{2}
\frac{1}{2|k_{1}^{0}|}\delta (k_{1}^{0}-k_{2}^{0})\sum_{s}s\Theta (sk_{1}^{0})\Theta (sk_{2}^{0})\delta({\bf k}_{1}-{\bf k}_{2})\sin [k_{1}\cdot(x_{2}-x_{1})]\nonumber\\
&&\cdot2k_{1}^{0}k_{1}^{\nu}
[W(x_{1})\Pi(x_{2})-\Pi(x_{1})W(x_{2})]\\
&=&\frac{1}{2(2\pi)^{4}}\int dx_{1}dx_{2} \int dk_{1}
\frac{\partial}{\partial x_{1\nu}}\cos [k_{1}\cdot(x_{1}-x_{2})] 
[W(x_{1})\Pi(x_{2})-\Pi(x_{1})W(x_{2})]\\
&=&\frac{1}{2}\int dx_{1}\Bigg[\Bigg(-\frac{\partial}{\partial x_{1\nu}}W(x_{1})\Bigg)\Pi(x_{1})+\Bigg(\frac{\partial}{\partial x_{1\nu}}\Pi(x_{1})\Bigg)W(x_{1})\Bigg]
 \end{eqnarray}
\end{subequations}
which is the result (10b), so $P_{w}^{\nu}=\int d{\bf x}T_{w}^{0\nu}(x)$.  Thus, all three constraints are satisfied. 

	In obtaining (12a), the integral over ${\bf x}$ gives  a factor $\delta({\bf k}_{1}-{\bf k}_{2})$, so 
\[	   
\delta({\bf k}_{1}-{\bf k}_{2})\Theta (sk_{1}^{0})\Theta (sk_{2}^{0})\delta (k_{1}^{2}-k_{2}^{2})=\delta({\bf k}_{1}-{\bf k}_{2})\Theta (sk_{1}^{0}) \delta (k_{1}^{0}-k_{2}^{0})/2|k_{1}^{0}|.
 \]
  Thus $k_{1}=k_{2}$,  so the term  
 $\sim \eta^{\mu\nu}$ vanishes, and $[k_{1}^{0}k_{2}^{\nu}+k_{1}^{\nu}k_{2}^{0}]=2k_{1}^{0}k_{1}^{\nu}$.  

	In going from  (12a) to (12b), we particularly note use of the identity $s\Theta (sk_{1}^{0} )/|k_{1}^{0}|=\Theta (sk_{1}^{0} )/k_{1}^{0}$.
	
	In going from  (12b) to (12c), we first integrate by parts, moving the derivative onto the operators, then integrate over $k_{1}$ which results in $\delta(x_{1}-x_{2})$, and then integrate over $x_{2}$.
	
	The stress tensor expression (11b) contains the form factor  
 \begin{equation}\label{13}
G(x-x_{1}, x-x_{2})\equiv\frac{1}{(2\pi)^{7}}\int dk_{1} dk_{2}
\delta (k_{1}^{2}-k_{2}^{2})\sum_{s}s\Theta (sk_{1}^{0})\Theta (sk_{2}^{0})\sin [k_{1}\cdot(x-x_{1})-k_{2}\cdot(x-x_{2})].
 \end{equation}
The stress tensor is not relativistic, because of the $s$-dependent terms in (\ref{13}), nor is it a local expression.  However, (\ref{13}) {\it is} responsible for the relativistic and local nature of the four-momentum,  since, following the steps outlined for Eqs.(\ref{12}),  
\begin{eqnarray}\label{14}
\int d {\bf x}G(x-x_{1}, x-x_{2})&=&\frac{1}{(2\pi)^{4}}\int dk_{1} dk_{2}
\frac{1}{|2k_{1}^{0}|}\delta (k_{1}^{0}-k_{2}^{0})\sum_{s}s\Theta (sk_{1}^{0})\Theta (sk_{2}^{0})\delta ({\bf k}_{1}-{\bf k}_{2})\sin [-k_{1}\cdot(x_{1}-x_{2})]\nonumber\\
&=&\delta ({\bf x}_{1}-{\bf x}_{2})\frac{1}{(2\pi)}\int d k_{1}^{0}\frac{1}{2k_{1}^{0}}\sin [k_{1}^{0}(t_{1}-t_{2})]=\frac{1}{4}\delta ({\bf x}_{1}-{\bf x}_{2})\epsilon(t_{1}-t_{2}).  
\end{eqnarray}
i.e., the time derivative of  (\ref{14}) is the local Lorentz scalar.  

	$G$ is calculated in Appendix B:
\begin{equation}\label{15}
G(x-x_{1}, x-x_{2})=-\frac{2}{(2\pi)^{4}}\frac{\partial^{2}}{\partial\sigma^{2}}\Bigg\{\Bigg[
\frac{T_{1}\Theta (\sigma+T_{1}^{2})}{\sqrt{\sigma+T_{1}^{2}}}+\frac{T_{2}\Theta (-\sigma+T_{2}^{2})}{\sqrt{-\sigma+T_{2}^{2}}}\Bigg]{\cal P}\frac{1}{\sigma}\Bigg\}
\end{equation}
where $\sigma\equiv (x-x_{1})^{2}- (x-x_{2})^{2}$, $T_{i}\equiv t-t_{i}$, and ${\cal P}$ denotes taking the principal part.  Its non-locality is evident in the location of the events $x_{1}$, $x_{2}$ where $\sigma$ vanishes, whose neighborhoods make the largest contribution to $G$ at the event $x$.  If one constructs  the forward and backward light-cones at $x$, and considers the family of (hyper-) hyperboloids of two sheets within these cones and asymptotically tangent to them, as well as the family of hyperboloids of one sheet outside these cones and tangent to them, $\sigma$ vanishes when both $x_{1}$ and $x_{2}$ lie on the same hyperboloid.  However,  in the application we shall be considering, this non-locality is severely truncated by considerations of what is \textit {probable}, as opposed to the above adumbration of what is \textit {possible}.

\section{CSL}
		
	The CSL Schr\"odinger-picture evolution of an initial state vector $|\phi\rangle$ is defined as 
\begin{equation}\label{16}
|\psi, t\rangle_{w}^{S}\equiv{\cal T}e^{-\int_{0}^{t}dt'\big[iH_{p}+\frac{1}{4\lambda}\int d{\bf x}' [ w({\bf x}',t') -2\lambda A({\bf x}')]^{2}\big]}|\phi\rangle.
\end{equation}
In (\ref{16}), ${\cal T}$ is the time-ordering operation, $H_{p}$ is the particle hamiltonian, and 
\begin{equation}\label{17}
A({\bf x})\equiv\frac{1}{(\pi a^{2})^{3/4}}\int d{\bf z}e^{-\frac{1}{2a^{2}}
({\bf z}-{\bf x})^{2}}\frac{1}{M_{0}}\sum_{n}M_{n}\xi_{n}^{\dagger}({\bf z})\xi_{n}({\bf z}),
\end{equation}
where the mass density operator (the sum in (\ref{17})) is expressed in terms of the creation and annihilation operators at 
${\bf x}$ of the nth type particle of mass $M_{n}$, and $M_{0}$ is the proton mass.  $\lambda$ and $a$ are respectively a collapse rate and a mesoscopic distance, parameters introduced by Ghirardi, Rimini and Weber\cite{GRW} for their collapse model, provisionally given the values $\lambda^{-1}\approx 10^{16}$sec, $a\approx 10^{-5}$cm (but, if the theory is proved correct, ultimately to be determined by experiment).  

The evolution (\ref{16}) is non-unitary, so the state-vector norm is not preserved.  CSL's rule for the probability density to be assigned to $w({\bf x},t)$ is 
\begin{equation}\label{18}
{\cal P}_{w}\equiv \negthinspace_{w}^{S}\langle \psi,t|\psi, t\rangle_{w}^{S}
\end{equation}
with $Dw\equiv \prod_{{\bf x},t}dw({\bf x},t)/\sqrt{2\pi\lambda /d{\bf x}dt}$ (rigorously definable in discrete space-time), so that $\int Dw{\cal P}_{w}=1$.  

Eqs.(\ref{16}-\ref{18}) completely define CSL.  With $H_{p}=0$,  if  states $|a_{j}\rangle$  describe different mass density distributions, it can be shown that an initial state $|\psi, 0\rangle=\sum_{j}c_{j}|a_{j}\rangle $  evolves into $\sim |a_{j}\rangle$ as $t\rightarrow\infty$, provided 
\[\lim_{t\rightarrow\infty}T^{-1}\int_{0}^{T} dt w({\bf x},t)\rightarrow      
2\lambda\int   \langle a_{j}|A({\bf x})|a_{j}\rangle.
\]
  Moreover,  it follows from (\ref{18}) that the probability measure of all such states is $|c_{j}|^{2}$. All fields with different asymptotic time averages have probability measure 0. With $H_{p}\neq 0$, this same behavior obtains provided the collapse dynamics is rapid compared to the Hamiltonian dynamics, which is usually the case.  

	Expressions simplify in the interaction picture, where the state vector only evolves due to its collapse dynamics.  Defining the interaction picture state 
vector $|\psi, t\rangle_{w}\equiv\exp{iH_{p}t}|\psi, t\rangle_{w}^{S}$,    (\ref{16}) becomes 
\begin{equation}\label{19}
|\psi, t\rangle_{w}\equiv{\cal T}e^{-\frac{1}{4\lambda}\int_{0}^{t}dt'\int d{\bf x}' [ w({\bf x}',t') -2\lambda A({\bf x}',t')]^{2}}|\phi\rangle.
\end{equation}
where $A({\bf x},t)\equiv\exp(iH_{p}t)A({\bf x})\exp(-iH_{p}t)$ (i.e., the particle creation and annihilation operators in 
(\ref{17}) become time dependent operators). The probability expression (\ref{18}) is unchanged, except that the $S$'s are removed.  It follows from this revised (\ref{18}) and (\ref{19}) that the density matrix which describes the ensemble of state vectors which evolve under all possible $w({\bf x},t)$ is
\begin{eqnarray}\label{20}
\rho(t)&=&\int Dw |\psi, t\rangle_{w}\negthinspace_{w}\langle \psi, t|\nonumber\\
&=&{\cal T}e^{-\frac{\lambda}{2}\int_{0}^{t}dt'\int d{\bf x}'[A_{L}({\bf x}',t')- A_{R}({\bf x}',t')]^{2}}\rho(0),
\end{eqnarray}
($\rho (0)=|\phi\rangle\langle\phi|$).  The first expression in (20) is so simple because the normalization factors which divide the un-normalized state vectors cancel the probability factor (\ref{18}). In (\ref{20}), the operators with subscript $L$ ($R$) appear to the left (right) of $\rho(0)$, and 
${\cal T}$ time-orders the operators to the left, and reverse-time-orders the operators to the right.  

\section{CQC and CSL}

The Schr\"odinger evolution of the state vector in CQC is 
\begin{equation}\label{21}
|\Psi, t\rangle^{S}=e^{-it[H_{p}+P_{w}^{0}+2\lambda\int d{\bf x}A({\bf x})\Pi({\bf x},0)]}|0\rangle|\phi\rangle.
\end{equation}
where $P_{w}^{0}$ is given by (10d), $\Pi({\bf x},0)$ is given by (\ref{3}), and $\langle w|0\rangle$ is given by (\ref{7}).
The CQC interaction picture state vector is $|\Psi, t\rangle=\exp it[H+P_{w}^{0}]|\Psi, t\rangle^{S}$, which yields
\begin{eqnarray}\label{22}
\langle w|\Psi, t\rangle&=&{\cal T}e^{-i2\lambda\int _{0}^{t}dt\int d{\bf x}
A({\bf x} ,t)(\delta/i\delta w({\bf x},t))}e^{-\frac{1}{4\lambda}\int_{-\infty}^{\infty}dt'
\int d{\bf x}'w^{2}({\bf x}',t')}|\phi\rangle\nonumber\\
&=&{\cal T}e^{-\frac{1}{4\lambda}\int_{-\infty}^{\infty}dt' 
\int d{\bf x}'[w({\bf x}',t')-2\lambda A({\bf x}' ,t')\Theta(t-t')\Theta(t')]^{2}}|\phi\rangle
\end{eqnarray}
using (\ref{7}).  One way to think of this is to regard $W({\bf x}, t)$ as a 
``pointer"  located at ${\bf x}$, labeled by $t$ (so there is an infinitude of pointers at ${\bf x}$).  The wave function for this pointer has an initial  (broad!) gaussian distribution $\exp-(1/4\lambda)dtd{\bf x}w^{2}({\bf x}, t)$ of possible ``positions" $w({\bf x}, t)$.  It evolves only during the time interval $(t, t+dt)$ to $\exp-(1/4\lambda)dtd{\bf x}[w({\bf x}, t)-2\lambda A({\bf x},t)]^{2}$, describing a standard (inaccurate!) local measurement of $A({\bf x},t)$.  
 
	The CQC state vector (\ref{22}) is identical to CSL's (\ref{19}), except for an extra numerical factor $f(t)$:
\[\langle w|\Psi, t\rangle= f(t)|\psi,t\rangle_{w} \hbox{\quad where\quad}     f(t)\equiv e^{-\frac{1}{4\lambda}\int_{-\infty}^{0}dt'
\int d{\bf x}'w^{2}({\bf x}',t')}e^{-\frac{1}{4\lambda}\int_{t}^{\infty}dt'
\int d{\bf x}'w^{2}({\bf x}',t')}.  
\]
 $f(t)$ does not play a role in calculations. It describes the wave functions for inessential pointers which either never will make measurements ($t'<0$) or which have yet to make measurements ($t'>t$).  Following the usual quantum mechanical rule, the  probability density associated with the eigenvalue $w({\bf x},t)$ of $W(x)$ is
$|\langle w|\Psi, t\rangle|^{2}$ which, when integrated over the inessential values of $w({\bf x}',t')$, is identical to the CSL probability rule (\ref{18}).  So, CSL and CQC 
are completely equivalent when it comes to making predictions concerning particles.

	However, CQC  does more.  It naturally provides an expression  for the energy $P_{w}^{0}$ of the white-noise field, as well as a particle-field interaction energy, such that the total energy $H\equiv H_{p}+P_{w}^{0}+2\lambda\int d{\bf x}' A({\bf x}')\Pi({\bf x}',0)$ is conserved.  And, as we have now seen, it allows definition of a stress tensor associated to the white-noise field. 
	
	The reason these structures have appeared in CQC, and not in CSL, can be seen by considering the CQC expectation value of any hermitian operator functional $F\{W(x), \Pi(x)\}$ which is polynomial in its arguments (but which may depend upon particle operators in any way), where the time argument of  $x$ lies in the range $(0,t)$, and then expressing it in terms of the CSL state vector:  
\begin{eqnarray}\label{23}
\langle \Psi,t|F\{W(x), \Pi(x)\} |\Psi, t\rangle&=&\int Dw\langle \Psi,t|w\rangle\langle w|F\{W(x), \Pi(x)\}|\Psi, t\rangle\nonumber\\
&&\negthinspace\negthinspace\negthinspace\negthinspace\negthinspace\negthinspace\negthinspace\negthinspace\negthinspace\negthinspace\negthinspace\negthinspace\negthinspace\negthinspace
\negthinspace\negthinspace\negthinspace\negthinspace\negthinspace\negthinspace\negthinspace\negthinspace\negthinspace\negthinspace\negthinspace\negthinspace\negthinspace\negthinspace
=\int Dw\langle \Psi,t|w\rangle F\{w(x), \frac{\delta}{i\delta w(x)}\}] \langle w|\Psi, t\rangle=
\int Dw_{w}\langle \psi,t|F\{w(x), \frac{\delta}{i\delta w(x)}\}|\psi, t\rangle_{w}.  
\end{eqnarray}
The functional derivative which appears in this expression means that $F$ is not a quantity which can be attributed to a single CSL state vector $|\psi, t\rangle_{w}$.  $F$'s expectation value depends upon a set of CSL state vectors infinitesimally close to $|\psi, t\rangle_{w}$.  At first glance, this appears to conflict with the accepted ontology of CSL (which is perhaps why no one looked for such expressions).  That is, reality is described by a single state vector.   Nature chooses a single $w(x)$ to evolve the state vector which is actually realized, and all other $w(x)$'s represent possible but unrealized evolutions.  

	However, two considerations may stimulate allowing such expressions within this ontology. One is that, for  state vectors with infinitesimally close $w(x)$'s, any difference in their description of  nature is indiscernible.  The other is that CSL already has consideration of an ensemble of infinitesimally close $w(x)$'s fundamentally built into it.  According to (\ref{18}),  $Dw_{w}\langle \psi,t|\psi, t\rangle_{w}$ is the probability to be associated with an ensemble of state vectors which have evolved under fields in an infinitesimal neighborhood of  $w(x)$. Thus, one may think of the family of state vectors infinitesimally close to the realized state vector as belonging to that state vector. 
	
	Therefore, consider  the CSL expression
\begin{equation}\label{24}
\frac{\negthinspace_{w}\langle \psi,t|F\{w(x), \frac{\overrightarrow\delta}{i\delta w(x)}\}|\psi, t\rangle_{w}+
\negthinspace_{w}\langle \psi,t|F\{w(x),\frac{\overleftarrow\delta}{-i\delta w(x)}\}]|\psi, t\rangle_{w}}{2\negthinspace_{w}\langle \psi,t|\psi, t\rangle_{w}}.
\end{equation}
(the arrows show the direction in which the functional derivative operates). This is real and,  when it is multiplied by the probability $Dw_{w}\langle \psi,t|\psi, t\rangle_{w}$ and $w$ is integrated over, with use of integration by parts and hermiticity of $F$, the result is  the ensemble expectation value (\ref{23}).  According to the argument given here, (\ref{24}) may be regarded as the expectation value of the operator $F$ for the CSL state $|\psi, t\rangle_{w}$. 

In the remainder of this paper, we shall discuss the ensemble-average energy-momentum density  in CSL, utilizing the CQC formalism to do so. 
	
\section{Ensemble-Average Energy-Momentum Density for Particles} 

 	Consider particle operators in the Heisenberg picture.   The annihilation operator for a particle of mass $m$ is $\xi(x)\equiv \xi({\bf x},t)=\exp (iH_{p}t)\xi({\bf x})\exp(-iH_{p}t)$.  For free non-relativistic particles,  where $H_{p}=(2m)^{-1}\int d{\bf x}{\bf \nabla}\xi^{\dagger}(x){\bf \cdot}{\bf \nabla}\xi(x)$, there is not an energy-momentum-stress density tensor.  There \textit{is} a mass density tensor $T_{m}^{\mu\nu}(x)$,   
\begin{eqnarray}\label{25}
T_{m}^{00}(x)&=&m\xi^{\dagger}(x)\xi(x), \quad T_{m}^{0i}(x)=T_{m}^{i0}(x)=\frac{1}{2i}\{
-[\partial_{i}\xi^{\dagger}(x)]\xi(x)+\xi^{\dagger}(x)[\partial_{i}\xi(x)]\}\nonumber\\ T_{m}^{ij}(x)&=&
(4m)^{-1}\{-[\partial_{i}\xi^{\dagger}(x)][\partial_{j}\xi(x)]-[\partial_{j}\xi^{\dagger}(x)][\partial_{i}\xi(x)]+[\partial_{i}\partial_{j}\xi^{\dagger}(x)] \xi(x)+ \xi^{\dagger}(x)    [\partial_{i}\partial_{j}\xi(x)] \}
\end{eqnarray}
whose $T_{m}^{0i}$ component (mass-density flux) is the momentum density, and there  \textit{is} an energy density tensor
$T_{e}^{\mu\nu}(x)$,
\begin{eqnarray}\label{26}
T_{e}^{00}(x)&=&(2m)^{-1}{\bf\nabla}\xi^{\dagger}(x){\bf\cdot}{\bf\nabla}\xi(x), \quad T_{e}^{0i}(x)=T_{e}^{i0}(x)=\frac{i}{2m}\{
-[\partial_{k}\xi^{\dagger}(x)]\partial_{k}\partial_{i}\xi(x)+\partial_{k}\partial_{i}\xi^{\dagger}(x)[\partial_{k}\xi(x)]\}\nonumber\\ 
T_{m}^{ij}(x)&=&
(2m)^{-2}\{-[\partial_{k}\xi^{\dagger}(x)][\partial_{k}\partial_{i}\partial_{j}\xi(x)]-[\partial_{k}\partial_{i}\partial_{j}\xi^{\dagger}(x)][\partial_{k}\xi(x)]\nonumber\\
&&\qquad\qquad+[\partial_{i}\partial_{k}\xi^{\dagger}(x)] [\partial_{j}\partial_{k}\xi(x)]+ [\partial_{j}\partial_{k}\xi^{\dagger}(x)] [\partial_{i}\partial_{k}\xi(x)] \}
\end{eqnarray}
whose $T_{e}^{00}$ component is the kinetic-energy density. The free particle tensors in Eqs.(\ref{25},\ref{26}) are conserved, $\partial_{\mu}T^{\mu\nu}=0$. [There  \textit{is} a conserved free particle relativistic energy-momentum density tensor, with  components 
\[T^{00}(x)=2^{-1}\Big\{\Big[\sqrt{-\nabla^{2}+m^{2}}\xi^{\dagger}(x)\Big]\xi(x)+\xi^{\dagger}(x)\Big[\sqrt{-\nabla^{2}+m^{2}}\xi(x)\Big]\Big\}, \quad T^{0i}(x)=T_{m}^{0i}(x),
\]
\[T^{ij}(x)=\frac{\sqrt{-\nabla_{1}^{2}+m^{2}}-\sqrt{-\nabla_{2}^{2}+m^{2}}}{\nabla_{1}^{2}-\nabla_{2}^{2}}
\Big[\xi^{\dagger}(x)\partial^{i}\partial^{j}\xi(x)+[\partial^{i}\partial^{j}\xi^{\dagger}(x)]\xi(x)-[\partial^{i}\xi^{\dagger}(x)]\partial^{j}\xi(x)-[\partial^{j}\xi^{\dagger}(x)]\partial^{i}\xi(x)\Big]
\]
($\nabla_{1}^{2}$ acts on $\xi^{\dagger}(x)$ and $\nabla_{2}^{2}$ acts on $\xi(x)$) which we shall not use here.]

	Suppose the particles interact through a potential, so $H_{p}$ includes the potential energy term 	
\[
V_{p}\equiv 2^{-1}\int d{\bf x}_{1}d{\bf x}_{2}\xi^{\dagger}(x_{1})\xi^{\dagger}(x_{2})V({\bf x}_{1}-{\bf x}_{2})\xi(x_{1})\xi(x_{2}).  
\]
Since the potential is non-local, one cannot define a local energy density.  One may define an energy density which depends on non-local contributions, 
 \[
 {\cal H}_{p}(x)\equiv T_{e}^{00}(x)+2^{-1}\xi^{\dagger}(x)\Bigg[\int d{\bf x}_{1}\xi^{\dagger}(x_{1})V({\bf x}_{1}-{\bf x})\xi(x_{1})\Bigg]\xi(x),
\]
 which has the virtue that its spatial integral equals $H_{p}$. However, one cannot form a symmetric conserved tensor with  ${\cal H}_{p}(x)$ as its 00th component. The mass-density tensor is also not conserved when there is an interaction:  	
\[\partial_{\nu}T_{m}^{\mu\nu}(x)=[1-\delta_{\mu 0}]\xi^{\dagger}(x)\Bigg[\partial^{\mu}\int d{\bf x}_{1}
V({\bf x}-{\bf x}_{1})\xi^{\dagger}(x_{1})\xi(x_{1})\Bigg]\xi(x),
\]
i.e., the divergence is the force density.  Nonetheless, we shall define the energy-momentum 	density 
operator for the particles as ${\cal P}_{p}^{\mu}(x)\equiv({\cal H}_{p}(x), T_{m}^{0i}(x))$.

	Now, consider the particle energy-momentum density in CQC.  Its expectation value (equal to the CSL ensemble average) for an initial particle state $|\phi\rangle$ and the initial vacuum field state $|0\rangle$ is, with help of (21), 
\begin{eqnarray}\label{27}
\overline{\cal P}_{p}^{\mu}(x)&=&^{S}\langle\Psi,t|{\cal P}_{p}^{\mu}({\bf x}, 0)|\Psi, t\rangle^{S}\nonumber\\
&=&\langle \phi |\langle 0|e^{it[H_{p}+P_{w}^{0}+2\lambda\int d{\bf x}A({\bf x})\Pi({\bf x},0)]}{\cal P}_{p}^{\mu}({\bf x}, 0)e^{-it[H_{p}+P_{w}^{0}+2\lambda\int d{\bf x}A({\bf x})\Pi({\bf x},0)]}|0\rangle|\phi\rangle\nonumber\\
&=&\langle \phi |\langle 0|{\cal T}e^{i2\lambda\int_{0}^{t} dx'A(x')\Pi(x')}{\cal P}_{p}^{\mu}(x)e^{-i2\lambda\int_{0}^{t} dx'A(x')\Pi(x')}|0\rangle|\phi\rangle
\end{eqnarray}
We note that (27) is expressed in terms of Heisenberg operators for the particles.  We now continue, by inserting $\int Dw |w\rangle\langle w|$ into (27), and utilizing (22):
\begin{equation}\label{28}
\overline{\cal P}_{p}^{\mu}(x)=\langle \phi |
{\cal T}\int Dw e^{-\frac{1}{4\lambda}\int_{0}^{t} 
dx'[w(x')-2\lambda A(x')]^{2}}{\cal P}_{p}^{\mu}(x)e^{-\frac{1}{4\lambda}\int_{0}^{t} 
dx'[w(x')-2\lambda A(x')]^{2}}|\phi\rangle
\end{equation}
 Writing
 \begin{equation}\label{29}
{\cal T}e^{-\frac{1}{4\lambda}\int_{0}^{t} 
dx'[w(x')-2\lambda A_{l}(x')]^{2}}e^{-\frac{1}{4\lambda}\int_{0}^{t} 
dx'[w(x')-2\lambda A_{r}(x')]^{2}}={\cal T}e^{-\frac{1}{2\lambda}\int_{0}^{t} 
dx'[w(x')-\lambda (A_{l}(x')+A_{r}(x'))]^{2}}e^{-\frac{\lambda}{2}\int_{0}^{t} 
dx'[A_{l}(x')-A_{r}(x')]^{2}}, 
\end{equation}
where the subscripts $l,r$ mean that the operators are respectively to the left and right of ${\cal P}_{p}^{\mu}(x)$, we can perform the integral over $w$ in (\ref{28}) and, with help of  (\ref{20}), we obtain 
\begin{eqnarray}\label{30}
\overline{\cal P}_{p}^{\mu}(x)&=&\langle \phi |
{\cal T}e^{-\frac{\lambda}{2}\int_{0}^{t} 
dx'[(A_{l}(x')-A_{r}(x')]^{2}}{\cal P}_{p}^{\mu}(x)|\phi\rangle\nonumber\\ 
&=&\hbox{Tr} {\cal P}_{p}^{\mu}(x){\cal T}
e^{-\frac{\lambda}{2}\int_{0}^{t}dx' [A_{L}(x')-A_{R}(x')]^{2} }
|\phi\rangle\langle\phi|=
\hbox{Tr}{\cal P}_{p}^{\mu}(x)\rho(t).
\end{eqnarray}
In  (\ref{30}), Tr  is the trace operation and the subscripts $L,R$ mean that the operators are respectively to the left and right of $|\phi\rangle\langle\phi|=\rho(0)$.

	One can proceed further by taking the time derivative of (30).  Employing ${\cal P}_{p}^{\mu}(x)=\exp iH_{p}t{\cal P}_{p}^{\mu}({\bf x},0)\exp -iH_{p}t$ and (\ref{17}), we have 
\begin{eqnarray}\label{31}
\frac{\partial}{\partial t}\overline{\cal P}_{p}^{\mu}(x)&=&-i\hbox{Tr}[{\cal P}_{p}^{\mu}(x),H_{p}]\rho(t)
-\frac{\lambda}{2}\int d{\bf x} \hbox{Tr}[A({\bf x},t),[A({\bf x},t),{\cal P}_{p}^{\mu}(x)]\rho(t)\nonumber\\
&=&-i\hbox{Tr}[{\cal P}_{p}^{\mu}(x),H_{p}]\rho(t)
-\frac{\lambda}{2}\int d{\bf z}  d{\bf z}' e^{-\frac{1}{4a^{2}}({\bf z}- {\bf z}')^{2} } \hbox{Tr}[\xi^{\dagger}({\bf z},t)\xi({\bf z},t),[\xi^{\dagger}({\bf z}',t)\xi({\bf z}',t),{\cal P}_{p}^{\mu}(x)]\rho(t).  
\end{eqnarray}
The double commutator is easily evaluated.  When $\mu=i$, the first commutator is $\sim\xi^{\dagger}(x)\xi(x)$, so the double commutator vanishes.  When $\mu=0$, the first commutator with $V_{p}(x)$ vanishes, so what is left is the double commutator with the kinetic energy density $T_{e}^{00}(x)$.  This commutator is 
\[ 
[\xi^{\dagger}({\bf z},t)\xi({\bf z},t),[\xi^{\dagger}({\bf z}',t)\xi({\bf z}',t),\frac{1}{2m}{\bf \nabla}
\xi^{\dagger}(x){\bf\cdot}{\bf \nabla}\xi(x]]=-\frac{1}{m}\xi^{\dagger}(x)\xi(x){\bf \nabla}_{z}\delta({\bf z}-{\bf x}){\bf\cdot}{\bf \nabla}_{z'}\delta({\bf z}'-{\bf x}).
\]
Putting this into (\ref{31}) yields the result 
\begin{eqnarray}\label{32}
\frac{\partial}{\partial t}\overline{\cal P}_{p}^{\mu}(x)&=&-i\hbox{Tr}[{\cal P}_{p}^{\mu}(x),H_{p}]\rho(t)-\delta^{\mu 0}\frac{\lambda}{2m}\int d{\bf z}  d{\bf z}' \delta({\bf z}-{\bf x})\delta({\bf z}'-{\bf x}){\bf \nabla}_{z}{\bf\cdot}{\bf \nabla}_{z'}e^{-\frac{1}{4a^{2}}({\bf z}- {\bf z}')^{2} } \hbox{Tr}\xi^{\dagger}(x)\xi(x)\rho(t)\nonumber\\
&=&-i\hbox{Tr}[{\cal P}_{p}^{\mu}(x),H_{p}]\rho(t)+\delta^{\mu 0}\frac{3\lambda}{4ma^{2}} \hbox{Tr}\xi^{\dagger}(x)\xi(x)\rho(t)
\end{eqnarray}
Thus, the ensemble-average energy density increase is strictly local, proportional to the expectation value of the particle number density.  

We note, when (\ref{32}) is integrated over ${\bf x}$, one obtains the well-known result that the particle energy increases linearly with time:
\[\frac{d}{dt}\overline H_{p}=\frac{3\lambda N}{4ma^{2}} 
\]
where $N$ is the number of particles.

\section{Ensemble-Average Energy-Momentum Density for $W$-Field} 

	Because of the non-local nature of $A({\bf x})$ given in (\ref{17}), the local-appearing CQC interaction (\ref{21})   is in fact non-local.  Thus, as for the particle tensors,  the $W$-field stress tensor is no longer conserved.  A calculation of the expectation value of the $W$-field stress tensor (equal to the CSL ensemble average)  
begins by using the expression (11a), 
\begin{subequations}\label{33} 
\begin{eqnarray}
T_{w}^{\mu\nu}(x)&=&\int dx_{1}dx_{2}\frac{1}{2}[W(x_{1})\Pi(x_{2})-W(x_{2})\Pi(x_{1})]G^{\mu\nu}(x-x_{1}, x-x_{2})\\
G^{\mu\nu}(x-x_{1}, x-x_{2})&\equiv& \frac{1}{(2\pi)^{7}} \int dk_{1} dk_{2}
\delta (k_{1}^{2}-k_{2}^{2})\sum_{s=-1}^{1}s\Theta (sk_{1}^{0})\Theta (sk_{2}^{0})\sin [k_{1}\cdot(x-x_{1})-k_{2}\cdot(x-x_{2})]\nonumber\\
&&\qquad\qquad\qquad\cdot[k_{1}^{\mu}k_{2}^{\nu}+k_{1}^{\nu}k_{2}^{\mu}+(1/2)\eta^{\mu\nu}(k_{1}-k_{2})^{2}].  
\end{eqnarray}
\end{subequations}
The expectation value of (33a) is  
\begin{eqnarray}\label{34}
\overline{T}_{w}^{\mu\nu}(x)&\equiv&^{S}\langle\Psi,t|T_{w}^{\mu\nu}({\bf x}, 0)|\Psi, t\rangle^{S}=\int dx_{1}dx_{2}\langle \phi |
{\cal T}\int Dw e^{-\frac{1}{4\lambda}\int_{0}^{t} 
dx'[w(x')-2\lambda A(x')]^{2}}\nonumber\\
&&\qquad\cdot \frac{1}{2}\Big[w(x_{1})\frac{\delta}{i\delta w(x_{2}})-w(x_{2})\frac{\delta}{i\delta w(x_{1}})\Big]e^{-\frac{1}{4\lambda}\int_{0}^{t} 
dx'[w(x')-2\lambda A(x')]^{2}}|\phi\rangle G^{\mu\nu}(x-x_{1}, x-x_{2})\nonumber\\
&=&\frac{-i}{2}\int_{0}^{t} dx_{1}dx_{2}\langle \phi |
{\cal T}\int Dwe^{-\frac{1}{2\lambda}\int_{0}^{t} 
dx'[w(x')-\lambda (A_{l}(x')+A_{r}(x'))]^{2}}e^{-\frac{\lambda}{2}\int_{0}^{t} 
dx'[A_{l}(x')-A_{r}(x')]^{2}}\nonumber\\
&&\qquad\cdot [w(x_{1})A_{r}(x_{2})-w(x_{2})A_{r}(x_{1})]|\phi\rangle G^{\mu\nu}(x-x_{1}, x-x_{2})
\end{eqnarray}
where we have used (\ref{29}) in the second step. Upon integration over $w$, the result is 
\begin{equation}\label{35}
\overline{T}_{w}^{\mu\nu}(x)=\frac{-i\lambda}{2}\int_{0}^{t} dx_{1}dx_{2}\langle \phi |
{\cal T}e^{-\frac{\lambda}{2}\int_{0}^{t} 
dx'[A_{l}(x')-A_{r}(x')]^{2}}
[A_{l}(x_{1})A_{r}(x_{2})-A_{l}(x_{2})A_{r}(x_{1})]|\phi\rangle G^{\mu\nu}(x-x_{1}, x-x_{2}).  
\end{equation}

	Eq. (\ref{35}) is complicated by the time ordering which embeds $A_{l}$, $A_{r}$ within the argument of the exponential.  However, a simplifying approximation can be made if the particle energies are much larger than 
$a^{-1}\approx$2eV.   When the integrals in (\ref{35}) over ${\bf x}_{1}$, ${\bf x}_{2}$ are performed, they act on the gaussians in 
 $A_{l}(x_{i})$, $A_{r}(x_{i})$ and on the argument of the sine in $G^{\mu\nu}$, giving
 \[\int d{\bf x}_{1}d{\bf x}_{2}e^{i{\bf k}_{1}{\bf \cdot}({\bf x}-{\bf x}_{1})-i{\bf k}_{2}{\bf \cdot}({\bf x}-{\bf x}_{2})
 }e^{-\frac{1}{2a^{2}}({\bf z}-{\bf x}_{1})^{2}}e^{-\frac{1}{2a^{2}}({\bf z}'-{\bf x}_{2})^{2}}\sim
 e^{i{\bf k}_{1}{\bf \cdot}({\bf z}-{\bf x}_{1})-i{\bf k}_{2}{\bf \cdot}({\bf z}'-{\bf x}_{2})}e^{-\frac{a^{2}}{2}[{\bf k}_{1}^{2}+{\bf k}_{2}^{2}]}.
 \]
That is, ${\bf k}_{j}^{2}\leq a^{-2}$, approximately.  Therefore, we make the approximation, in the expression (33b) for $G^{\mu\nu}$,
\[
\delta(k_{1}^{2}-k_{2}^{2})=\delta(k_{1}^{02}-{\bf k}_{1}^{2}-k_{2}^{02}+{\bf k}_{2}^{2})\approx \delta(k_{1}^{02}-k_{2}^{02}).
\] 
With this approximation, the form factor $G$ in (\ref{13}) and $G^{0\nu}$ become
\begin{subequations}\label{36}
\begin{eqnarray}
G(x-x_{1}, x-x_{2})&\approx&\frac{1}{4} \delta({\bf x}-{\bf x}_{1})\delta({\bf x}-{\bf x}_{2})\epsilon(t_{1}-t_{2}),\\
 G^{0\nu}(x-x_{1}, x-x_{2})&\approx&[\partial_{1}^{\nu}-\partial_{2}^{\nu}]\frac{1}{2} \delta({\bf x}-{\bf x}_{1})\delta({\bf x}-{\bf x}_{2})\delta(t_{1}-t_{2}).
\end{eqnarray}
\end{subequations}

	If we put (36b) into (33a), there results 
\begin{equation}\label{37}
T_{w}^{0\nu}(x)\approx\int dt\frac{1}{2}\big[W({\bf x},t)\frac{\partial}{\partial x_{\nu}}\Pi({\bf x},t)-\Pi({\bf x},t)\frac{\partial}{\partial x_{\nu}}W({\bf x},t)\Big].
\end{equation}
This (time-independent) expression for the energy-momentum density is what we would have obtained if we had looked at the 
 the $W$-field energy-momentum $P_{w}^{\nu}$ in (10b), and just removed the integral over ${\bf x}$.
 
	If we put (36b) into (\ref{35}) there results
\begin{subequations}\label{38}
\begin{eqnarray}
\overline{T}_{w}^{0\nu}(x)&\approx&\frac{-i\lambda}{2}\int_{0}^{t} dt_{1}\langle \phi |
{\cal T}e^{-\frac{\lambda}{2}\int_{0}^{t_{1}} 
dx'[A_{l}(x')-A_{r}(x')]^{2}}
[A({\bf x},t_{1}),\partial^{\nu}A({\bf x},t_{1})]|\phi\rangle\label{38a}\\
&=&\frac{\lambda}{2}\int_{0}^{t} dt_{1}
\hbox {Tr}
[A({\bf x},t_{1}),[A({\bf x},t_{1}),P_{w}^{\nu}]]\rho(t_{1}).\label{38b}
\end{eqnarray}
\end{subequations}
In (\ref{38a}), the exponential's integral's upper limit has been changed from $t$ to $t_{1}$ since the time-ordered integral from $t_{1}$ to $t$ vanishes.  The subscripts $l,r$ have been dropped from the commutator since they are evaluated at the same time, and are no longer buried within the time ordering of the exponential's argument.  We note that, when (\ref{38b}) is integrated over ${\bf x}$ to obtain the $W$-field energy-momentum, and the time derivative is taken, the result is the negative of the ${\bf x}$-integral of (\ref{31}) so the total energy-momentum of particles and $W$-field  is conserved (the interaction term makes no contribution to the ensemble-average energy).    

	As in the previous section, one can proceed further by evaluating the double commutator in (\ref{38b}), obtaining   
\begin{equation}\label{39}
\overline{T}_{w}^{0\nu}(x)\approx\delta^{\nu 0}\frac{-\lambda}{2\pi^{3/2}ma^{7}}\int d{\bf z}({\bf z}-{\bf x})^{2}
e^{-\frac{1}{a^{2}}({\bf z}-{\bf x})^{2}}\int_{0}^{t}dt_{1}\hbox{Tr}\xi^{\dagger}({\bf z},t_{1})\xi({\bf z},t_{1})\rho(t_{1}).
\end{equation}

From (\ref{39}) we see that the ensemble-average energy density is negative. It is also non-local, acquiring an increment at ${\bf x}$, during  $(t_{1},t_{1}+dt_{1}) $,  proportional to the particle number density at time $t_{1}$  in a volume $\approx a^{3}$ around ${\bf x}$.  

	We also note that the energy in the $W$-field, according to (\ref{39}), is
\[
P_{w}^{0}(t)=\int d{\bf x} \overline{T}_{w}^{0\nu}(x)=-\frac{3\lambda Nt}{4ma^{2}} ,
\]
with rate of change equal to the negative of the particle energy rate of change given in the equation following (\ref{32}).

\section{Particle Creation Model} 
	
	From the last equation in the last section, according to CSL, over the age of the universe, a proton (mass m) gains energy and the  $W$-field loses energy $\approx 10^{-16}mc^{2}$.  This is too small to have any cosmological significance.  
	
	However, one may speculate that collapse could play a significant role in the creation of the universe\cite{iceland}.  If the universe obeys quantum theory in its initial stages, it is likely that the Hamiltonian which governs it causes its state vector to evolve into a superposition of possible universes.  The choice of which universe is actually ours could be due to a collapse mechanism.  Suppose that CSL  provides this mechanism.  A significant amount of $W$-field energy density could be produced, whose gravitational influence could thereafter have a role to play in the further evolution of the universe, for example, in its expansion or in galactic formation.    

	A simple model of particle creation to illustrate the generation of such $W$-field energy has been presented\cite{iceland}.   Here, we shall reconsider that model, and calculate for it the particle and  $W$-field energy densities.  The particle energy density is
\begin{equation}\label{40}
{\cal H}_{p}(x)=m\xi^{\dagger}(x)\xi(x)+g({\bf x})[\xi^{\dagger}(x)+\xi(x)],   
\end{equation}
so the particle Hamiltonian is $H_{p}=\int d{\bf x}{\cal H}_{p}(x)$.  
This describes a ``displaced"  harmonic oscillator at each ${\bf x}$, 	
with displacement $\sim g({\bf x})$. From an initial no-particle state $|0\rangle_{p}$,  without collapse, the ensemble average of the particle number density 
${ \cal N}_{p}(x)=\xi^{\dagger}(x)\xi(x)$ just oscillates.  As we shall see, with collapse, it steadily grows as the oscillation dies out.   

	We first consider the ensemble-average particle energy density.  Eq.(\ref{31}) holds for any Heisenberg particle operator density, not just $\overline{\cal P}_{p}^{\mu}(x)$, so 
\begin{eqnarray}\label{41}
\frac{\partial}{\partial t}\overline{\cal N}_{p}(x)&=&-i\hbox{Tr}[{\cal N}_{p}(x),H_{p}]\rho(t)
-\frac{\lambda}{2}\int d{\bf x} \hbox{Tr}[A({\bf x},t),[A({\bf x},t),{\cal N}_{p}(x)]\rho(t)\nonumber\\
&=&-i\hbox{Tr}[{\cal N}_{p}(x),H_{p}]\rho(t)=-ig({\bf x})[\overline\xi^{\dagger}(x)-\overline\xi(x)],
\end{eqnarray}
\begin{equation}\label{42}
\frac{\partial}{\partial t}\overline\xi(x)=-i\hbox{Tr}[\xi(x),H_{p}]\rho(t)
-\frac{\lambda}{2}\int d{\bf x} \hbox{Tr}[A({\bf x},t),[A({\bf x},t),\xi(x)]\rho(t)=
-\Big[im+\frac{\lambda}{2}\Big]\overline\xi(x)-ig({\bf x}).
\end{equation}
The solution of (\ref{42}) is
\begin{equation}\label{43}
\overline\xi(x)=\frac{-i g({\bf x})}{im+(\lambda/2)}\big[1-e^{-[im+(\lambda/2)]t}\big].  
\end{equation}

Thus,  (\ref{41}) and (\ref{40}) give respectively
\begin{equation}\label{44}
\frac{\partial}{\partial t}\overline{\cal N}_{p}(x)=\frac{2g^{2}({\bf x})}{m^{2}+(\lambda/2)^{2}}\Big\{
\frac{\lambda}{2}\big[1-e^{-{\lambda t/2}}\cos mt\big]
+m e^{-\lambda t/2}\sin m t\Big\},   
\end{equation}
\begin{equation}\label{45}
\frac{\partial}{\partial t}\overline{\cal H}_{p}(x)=\frac{g^{2}({\bf x})\lambda}{m^{2}+(\lambda/2)^{2}}\Big\{
m\big[1-e^{-{\lambda t/2}}\cos mt\big]
-\frac{\lambda}{2} e^{-\lambda t/2}\sin m t\Big\}.   
\end{equation}
When there is no collapse ($\lambda=0$), we see that $\overline{\cal N}_{p}(x)$ just oscillates, and 
$\overline{\cal H}_{p}(x)$ remains constant (each oscillator at ${\bf x}$ conserves energy).  When $\lambda\neq 0$, $\overline{\cal N}_{p}(x)$ and $\overline{\cal H}_{p}(x)$ asymptotically grow linearly with time, with growth rate  $\sim g^{2}({\bf x})\lambda$. Thus,  the permanent creation of particles depends upon the collapse mechanism.  

	The $W$-field energy density can be found from (\ref{38b}):
\begin{eqnarray}\label{46}
\overline T_{w}^{00}(x)&\approx&\frac{\lambda}{2}\int_{0}^{t} dt_{1}
\hbox {Tr}[A({\bf x},t_{1}),[A({\bf x},t_{1}),H_{p}]]\rho(t_{1})\nonumber\\
&=&\frac{\lambda}{2(\pi a^{2})^{3/2}}\int_{0}^{t} dt_{1}\int d{\bf z}e^{-\frac{1}{a^{2}}({\bf x}-{\bf z})^{2}}g({\bf z})
[\overline\xi^{\dagger}({\bf z}, t_{1})+\overline\xi({\bf z}, t_{1})]\nonumber\\
&=&\frac{-\lambda}{2(\pi a^{2})^{3/2}[m^{2}+(\lambda/2)^{2}]}
\int d{\bf z}e^{-\frac{1}{a^{2}}({\bf x}-{\bf z})^{2}}g^{2}({\bf z})
\int_{0}^{t} dt_{1}\Big\{
m\big[1-e^{-{\lambda t_{1}/2}}\cos mt_{1}\big]
-\frac{\lambda}{2} e^{-\lambda t_{1}/2}\sin m t_{1}\Big\}. 
\end{eqnarray}
The salient features here are that this energy density is negative, it  is generated non-locally (the contribution at ${\bf x}$ comes from a distance $\approx a$ about ${\bf x}$), and it grows linearly with time for $t>>\lambda^{-1}$.  The total energy is conserved 
(the sum of the spatial integral of the time-derivative of  (\ref{46}) and the spatial integral of 
 (\ref{45}) equals 0).   
\section{W-Field Energy Density as a Gravitational Source}
	We shall consider an example in which the $W$-field energy density is a source of gravitation.   
The Hamiltonian which governs the state vector evolution in the CQC Schr\"odinger picture is taken to be 
\begin{equation}\label{47}	  
H\equiv \int dx T_{w}^{00}(x)\Bigg[1-GM\int d {\bf z}\frac{{\cal N}_{p}( {\bf z})}{| {\bf x}- {\bf z}|}\Bigg]+
M\int d{\bf x}{\cal N}_{p}( {\bf x})\Bigg[1-\frac{GM}{2}\int d {\bf z}\frac{{\cal N}_{p}( {\bf z})}{| {\bf x}- {\bf z}|}\Bigg]+
2\lambda \int d{\bf x}A({\bf x})\Pi({\bf x}, 0)
\end{equation}
In Eq. (\ref{47}), there are three terms: call them $A$, $B$ and $C$ respectively.  $A$ consists of the W-field energy and the gravitational energy of its interaction with particles of mass $M$.  For simplicity, we shall take $T_{w}^{00}(x)$ to be given by the approximate expression (\ref{37}).  $B$ consists of the particle mass-energy and the particle 
gravitational self-interaction energy.  $C$ is the usual CQC interaction  between the W-field and particles, given in (\ref{27}), which is responsible for the collapse dynamics, where  $A({\bf x})$ is given by (\ref{17}):
\begin{equation}\label{48}
A({\bf x})\equiv\frac{1}{(\pi a^{2})^{3/4}}\int d{\bf z}e^{-\frac{1}{2a^{2}}
({\bf z}-{\bf x})^{2}}\frac{1}{M_{0}}M{\cal N}_{p}( {\bf z}).
\end{equation}

	What is not present in (\ref{47}) is the particle kinetic and potential energy.  This is usual when one wishes to display collapse behavior without interference by particle dynamics.   It is a good approximation when the collapse time is shorter than the time scale over which the states evolve appreciably.  Also omitted is the W-field  gravitational self-interaction energy  and the gravitational contribution of the particle kinetic and potential energy-densities, 
which are presumed to be negligibly small.  

	In the CQC interaction picture, 	
\[
|\Psi, t\rangle=e^{it(A+B)}e^{-itH}|\Psi, 0\rangle, 
\]
from which follows 
\begin{eqnarray}\label{49} 
\frac{d}{dt}	|\Psi, t\rangle&=&e^{it(A+B)}(-iC)e^{-it(A+B)}|\Psi, t\rangle\nonumber\\
&=&e^{itA}(-iC)e^{-itA}|\Psi, t\rangle,
\end{eqnarray}	
since $B$ commutes with both $A$ and $C$.  	
We now expand the initial particle state in eigenstates $|n_{r}\rangle$ of the number-density operator 
${\cal N}_{p}({\bf z})$ (${\cal N}_{p}( {\bf z})|n_{r}\rangle=n_{r}( {\bf z})|n_{r}\rangle$), 
\[
|\Psi, 0\rangle=|0\rangle|\phi\rangle=|0\rangle\sum_{r}c_{r}(0)|n_{r}\rangle 
\]
(for definiteness, the number density eigenstates have been given discrete labels). Since ${\cal N}_{p}( {\bf z})$ is the only operator in the Hamiltonian which acts on the particle state vector, the states $|n_{r}\rangle$ do not alter during the state vector evolution. Thus the operator ${\cal N}_{p}( {\bf z})$ in $A$ and $C$ in (\ref{49}) may be replaced by the c-number $n_{r}( {\bf z})$. With that replacement,  $A({\bf x})$ given by (\ref{48}) will be denoted $A_{r}({\bf x})$.

To evaluate (\ref{49}), we need 
to calculate $\exp(itA)\Pi({\bf x},0)\exp(-itA)$ where 
\begin{eqnarray}\label{50}
A&=&\int dx'\frac{1}{2}[\dot W({\bf x}',t')\Pi({\bf x}',t')-\dot\Pi({\bf x}',t')W({\bf x}',t')]\Bigg[1-GM\int d{\bf z}
\frac{n_{r}( {\bf z})}{|{\bf x}'-{\bf z}|}\Bigg]\nonumber\\
&=& \int dx'[-\dot\Pi({\bf x}',t')W({\bf x}',t')][1+\phi_{r}({\bf x}')].  
\end{eqnarray}
It follows that 
\begin{eqnarray}\label{51}
e^{itA}\Pi({\bf x},0)e^{-itA}=\Pi({\bf x},0)+t[1+\phi_{r}({\bf x})]\dot\Pi({\bf x},0)+...=\Pi\big({\bf x},t[1+\phi_{r}({\bf x})]\big).
\end{eqnarray}
	Thus, the solution of (\ref{49}) is 
\begin{eqnarray}\label{52}
	|\Psi, t\rangle&=&\sum_{r}c_{r}(0)e^{-i2\lambda\int d{\bf x}'\int_{0}^{t}dt'\Pi({\bf x}',t'[1+\phi_{r}({\bf x}')])A_{r}({\bf x}')}|0\rangle|n_{r}\rangle\nonumber\\
	&=&\sum_{r}c_{r}(0)e^{-i2\lambda\int d{\bf x}\int_{0}^{t[1+\phi_{r}({\bf x})]}d\tau\Pi({\bf x},\tau ) A_{r}({\bf x})[1+\phi_{r}({\bf x})]^{-1}}|0\rangle|n_{r}\rangle.		
\end{eqnarray}	
In (\ref{52}) we have made a change of variables $\tau=t'[1+\phi_{r}({\bf x}')]$, ${\bf x}={\bf x}'$, so 
$dt'd{\bf x}'=d\tau d{\bf x}[1+\phi_{r}({\bf x})]^{-1}$.  

	We may now proceed, as in (\ref{22}), to go to the $|w\rangle$ basis and obtain an expansion of the state vector whose terms correspond to CSL state vectors: 
\begin{eqnarray}\label{53}	
	|\Psi, t\rangle&=&\sum_{r}c_{r}(0)|n_{r}\rangle\int Dw|w\rangle e^{-i2\lambda\int d{\bf x}\int_{0}^{t[1+\phi_{r}({\bf x})]}d\tau A_{r}({\bf x})[1+\phi_{r}({\bf x}')]^{-1}\delta/i\delta w({\bf x}, \tau)}e^{-\frac{1}{4\lambda}\int dx'w^{2}(x')}\nonumber\\
	&=&\sum_{r}c_{r}(0)|n_{r}\rangle\int Dw|w\rangle e^{-\frac{1}{4\lambda}\int d{\bf x}'\int_{-\infty}^{\infty}dt'\big[w({\bf x}',t')-A_{r}({\bf x}')[1+\phi_{r}({\bf x}')]^{-1}
	\Theta\big( t[1+\phi_{r}({\bf x}')]-t'\big)\Theta(t')\big]^{2}}\nonumber\\
	&=&\sum_{r}c_{r}(0)|n_{r}\rangle\int Dw|w\rangle f(t)e^{-\frac{1}{4\lambda}\int d{\bf x}'\int_{0}^{t[1+\phi_{r}({\bf x}')]}dt'\big[w({\bf x}',t')-A_{r}({\bf x}')[1+\phi_{r}({\bf x}')]^{-1}\big]^{2}}
\end{eqnarray}
 where
 \[
 f(t)=e^{-\frac{1}{4\lambda}\int d{\bf x}'\int_{-\infty}^{0}w^{2}({\bf x}',t')}e^{-\frac{1}{4\lambda}\int d{\bf x}'\int_{t[1+\phi_{r}({\bf x}')]}^{\infty}w^{2}({\bf x}',t')}.
 \] 

	Eq. (\ref{53}) is identical in form to Eq. (\ref{22}) except for two outstanding distinctions.  
	
	The first is that,  in (\ref{22}),  $|\Psi, t\rangle$ is determined by  $w({\bf x}',t')$ values  between the hypersurfaces $t'=0$ and $t'=t$ for all ${\bf x}'$. However, in (\ref{53}), the relevant $w({\bf x}',t')$ values lie between the hypersurfaces $t'=0$  
and $t'=t[1+\phi_{r}({\bf x}')]$ for all $x'$. 

	Otherwise, the collapse behavior is the same.  The  CSL  states
\[
|\psi, t\rangle_{w}=\sum_{r}c_{r}(0)|n_{r}\rangle e^{-\frac{1}{4\lambda}\int d{\bf x}'\int_{0}^{t[1+\phi_{r}({\bf x}')]}dt'\big[w({\bf x}',t')-A_{r}({\bf x}')[1+\phi_{r}({\bf x}')]^{-1}\big]^{2}}
\]
can be plucked from  (\ref{53}) since, according to it,  if two $w({\bf x}',t')$ are identical between the above specified hypersurfaces but differ in the infinitesimal slice hypervolume  between $t'=(t[1+\phi_{r}({\bf x}')], (t+dt)[1+\phi_{r}({\bf x}')])$, the associated states are orthogonal forever after.  

Moreover, as is well known, if $t$ is the time read by a clock far from gravitational sources, the gravitational time dilation effect is such that, where there is a weak static gravitational potential, a local clock at ${\bf x}'$ reads time 
\[
t'=t\sqrt{-g_{00}({\bf x}')}= t\sqrt{1+2\phi({\bf x}')}\approx t[1+\phi({\bf x}')].  
\]	
That is, instead of collapse taking place on evolving $t$ hypersurfaces, the Hamiltonian (\ref{47}) describes collapse as taking place on evolving local proper time hypersurfaces.  

The second distinction is that Eq. (\ref{22}) describes collapse toward eigenstates of $A({\bf x})$, but  Eq. (\ref{53}) describes collapse toward eigenstates of 
$A({\bf x})[1+\phi({\bf x})]^{-1}$.   Actually, that has no essential effect on the collapse behavior, since $A$ and 
$\phi$ both depend only on  ${\cal N}_{p}( {\bf z})$, and so collapse in both cases is toward the eigenstates $|n_{r}\rangle$.  

However, there is a subtle difference. The  off-diagonal elements of the density matrix have different decay rates. In the first case, the decay of the $r-s$ element is proportional to 	$\int d{\bf x}(A_{r}({\bf x})-A_{s}({\bf x}))^{2}$, while in the second case it is proportional to 
\[
\int d{\bf x}\big(A_{r}({\bf x})[1+\phi_{r}({\bf x})]^{-1}-A_{s}({\bf x})[1+\phi_{r}({\bf x})]^{-1}\big)^{2}.  
\]	
So, the question arises as to how to understand this or, more broadly, how to interpret the appearance of $A_{r}({\bf x}')[1+\phi_{r}({\bf x}')]^{-1}$ in Eq. (\ref{53}).

Here is one point of view.  Suppose that the gravitational potential of a point mass is not the Newtonian value, but rather is  that due to the mass smeared by a gaussian over the scale $a$.  So far, there is no experimental restriction on this possibility, for $a\approx 10^{-5}$cm\cite{grav}.  Define 
\[
\tilde{\cal N}_{p}({\bf x})\equiv\int d{\bf z}\frac{1}{(2\pi a^{2})^{3/2}}
e^{-\frac{1}{2a^{2}} ({\bf x}-{\bf z})^{2}}{\cal N}_{p}({\bf z})
\]
so the gravitational potential is 
\[
\tilde{\phi}({\bf x})\equiv-Gm\int d{\bf z}   \frac{\tilde{\cal N}_{p}({\bf z})}{|{\bf x}-{\bf z}|}.  
\]

	According to (\ref{48}), 
\[
A({\bf x})=(4\pi a^{2})^{3/4}\frac{M}{M_{0}}\tilde{\cal N}_{p}({\bf x}).
\]
Now, it is not unreasonable to replace the particle's mass-energy density $\sim A({\bf x})$ by the particle's total energy density, mass-plus gravitational, in the term C in the Hamiltonian, i.e., 
\[	
A({\bf x})\rightarrow\tilde{A}({\bf x})\equiv A({\bf x})[1+\tilde{\phi}({\bf x})]. 
\]

	With the Hamiltonian (\ref{47}) replaced by 	
\begin{equation}\label{54}	  
H\equiv \int dx T_{w}^{00}(x)[1+\tilde{\phi}({\bf x})]+
M\int d{\bf x}\tilde{\cal N}_{p}( {\bf x})[1+\tilde{\phi}({\bf x})]+
2\lambda \int d{\bf x}A({\bf x})[1+\tilde{\phi}({\bf x})]\Pi({\bf x}, 0),
\end{equation}
the argument goes through as before, with the result 
\begin{eqnarray}\label{55}	
	|\Psi, t\rangle=\sum_{r}c_{r}(0)|n_{r}\rangle\int Dw|w\rangle f(t)e^{-\frac{1}{4\lambda}\int d{\bf x}'\int_{0}^{t[1+\tilde{\phi}_{r}({\bf x}')]}dt'[w({\bf x}',t')-A_{r}({\bf x}')]^{2}}.  
\end{eqnarray}
That is, we have put  the total energy density $\tilde{A}({\bf x})$ into the Hamiltonian's interaction term C, but the collapse dynamics is determined by ${A}({\bf x})$, the mass-energy density, as is usual in CSL.

\appendix
\section{Obtaining the Energy-Momentum-Stress Density Tensor}
Here, a method of  obtaining the stress tensor (\ref{11}) is presented.  The idea is to think of $W(x)$ as composed of free quantum scalar fields of all possible masses (bradyonic and tachyonic) and energies (positive and negative). Although  $W(x)$ does not have a Lagrangian, each such quantum field does have a Lagrangian and thereby a stress tensor, and when these stress tensors are added up, they yield a stress tensor for $W(x)$.

Define the quantum fields 
\begin{equation}\label{A1}
\phi_{m,r,s}(x)\equiv\frac{1}{(2\pi)^{3/2}}\int dk \delta (k^{2}+rm^{2})\Theta (sk^{0})
[b(k)e^{ik\cdot x}+b^{\dagger}(k)e^{-ik\cdot x}]
\end{equation}
In (\ref{A1}), $m$ is the mass of the field, $r=1$ or $-1$ if the field is respectively bradyonic or tachyonic, 
$s=1$ or $-1$ if the energy is respectively positive or negative.  One readily finds the commutator
\begin{eqnarray}\label{A2}
[\phi_{m,r,s}(x), \dot\phi_{m',r',s'}(x')]&=&\delta(m^{2}-m'^{2})\delta_{rr'}\delta_{ss'}\nonumber\\
&\cdot&\frac{i}{2(2\pi)^{3}}
\int d{\bf k}\Theta({\bf k}^{2}+rm^{2})[e^{i{\bf k}{\bf \cdot}({\bf x}-{\bf x}') -is\omega_{m,r}({\bf k})(t-t')}+
e^{-i{\bf k}{\bf\cdot}({\bf x}-{\bf x}')+ is\omega_{m,r}({\bf k})(t-t')}]
\end{eqnarray}
where $\omega_{m,r}({\bf k})\equiv\sqrt{{\bf k}^{2}+rm^{2}}$.  For fixed $m=m',r=r',s=s'$, this would be the usual commutation relation, except for the factor 
$\delta(m^{2}-m^{2})=\delta(0)$.  If heuristically we think of $\delta (0)=1/dm^{2}$,  we may think of 
$\sqrt{dm^{2}}\phi_{m,r,s}(x)$ as a usual quantum field.  Then, we may construct the usual stress tensor for these usual quantum fields, and add them up, getting a combined stress tensor for all possible four-momenta, 
\begin{eqnarray}\label{A3}
T_{w}^{\mu\nu}(x)&=&\sum_{rs}\int_{0}^{\infty}dm^{2}\frac{s}{2}\Big\{\partial^{\mu}\phi_{m,r,s}(x)\partial^{\nu}\phi_{m,r,s}(x)+\partial^{\nu}\phi_{m,r,s}(x)\partial^{\mu}\phi_{m,r,s}(x)\nonumber\\
&-&\eta^{\mu\nu}\big[\partial_{\lambda}\phi_{m,r,s}(x)\partial^{\lambda}\phi_{m,r,s}(x)+\frac{1}{2}\big(\phi_{m,r,s}(x)\partial_{\lambda}\partial^{\lambda}\phi_{m,r,s}(x)+\partial_{\lambda}\partial^{\lambda}\phi_{m,r,s}(x)\phi_{m,r,s}(x)\big)\big]\Big\}
\end{eqnarray}
In (\ref{A3}),  the usual mass-squared term $rm^{2}\phi_{m,r,s}^{2}(x)$ has been rewritten utilizing the dynamical equation
\[rm^{2}\phi_{m,r,s}(x)=\partial_{\lambda}\partial^{\lambda}\phi_{m,r,s}(x). 
\]
 The factor $s$   in (\ref{A3}) gives the negative energy fields their negative energy.  Note that relativistic invariance has been broken by treating the stress tensor for the positive and negative energy tachyon fields separately.  This is because relativistic tachyon quantum field theory  requires exchange of tachyon creation and annihilation operators when the tachyon energy changes sign\cite{Sudarshan}, but no such exchange takes place for $b(k)$,  $b^{\dagger}(k)$.  

	The four-divergence of this stress tensor vanishes when the dynamical equation is utilized:  
\begin{eqnarray}\label{A4}
\partial_{\nu}T_{w}^{\mu\nu}&=&\sum_{rs}\int_{0}^{\infty}dm^{2}\frac{s}{4}
\{[\partial^{\mu}\phi_{m,r,s}(x)][\partial_{\lambda}\partial^{\lambda}\phi_{m,r,s}(x)]
+[\partial_{\lambda}\partial^{\lambda}\phi_{m,r,s}(x)][\partial^{\mu}\phi_{m,r,s}(x)]\nonumber\\
&&\qquad\qquad\qquad-
[\partial^{\mu}\partial_{\lambda}\partial^{\lambda}\phi_{m,r,s}(x)]\phi_{m,r,s}(x)
-\phi_{m,r,s}(x)[\partial^{\mu}\partial_{\lambda}\partial^{\lambda}\phi_{m,r,s}(x)]\}=0.
\end{eqnarray}

That this stress tensor satisfactorily represents the $W$-field energy-momentum is shown by calculating the four-momentum which follows from (\ref{A3}), utilizing (\ref{A1}): 
\begin{subequations}
\begin{eqnarray}\label{A5} 
P_{w}^{\nu}=\int d {\bf x} T_{w}^{0\nu}(x)&=&\sum_{rs}\int_{0}^{\infty}dm^{2}s\int dkdk'\delta (k^{2}+rm^{2})\delta (k'^{2}+rm^{2})\Theta (sk^{0})\Theta (sk'^{0})\delta ({\bf k}-{\bf k}')2b^{\dagger}(k)b(k)\nonumber\\
&&\quad\quad\quad\quad\cdot\{k^{0}k'^{\nu}+\delta^{0\nu}(1/2)[k\cdot k'-k'^{2}]\}\\
&=&\sum_{rs}s\int dk \Theta (sk^{0})\Theta[r(|k^{0}|-|{\bf k}|)]\frac{1}{2|k^{0}|}
k^{0}k^{\nu}2b^{\dagger}(k)b(k)\\
&=&\int d{\bf k} \Bigg\{\int_{|{\bf k}|}^{\infty}+\int_{-\infty}^{-|{\bf k}|}+\int_{0}^{|{\bf k}|}+\int_{-|{\bf k}|}^{0}\Bigg\}       
dk^{0}k^{\nu}b^{\dagger}(k)b(k)=\int dkk^{\nu}b^{\dagger}(k)b(k), 
\end{eqnarray}
\end{subequations}	
which is identical to (\ref{8}).  

The term $\sim b(k)b(k')$ (also its Hermitian conjugate) is not present in (A5a), for the following reasons.  The delta function $\delta ({\bf k}+{\bf k}')$ and step functions make it $\sim b({\bf k},k^{0})b({-\bf k},k^{0})$. When $\nu\neq 0$, this is symmetric under change of sign and the curly bracketed factor ${\bf k}$ which multiplies it is antisymmetric so the integral over ${\bf k}$ vanishes.  When $\nu= 0$, the  curly bracketed factor which multiplies it is $\{-k^{0}k'^{0}+(1/2)[-k\cdot k'-k'^{2}]\}$, which vanishes for ${\bf k}=-{\bf k}'$ and 
$k^{0}=k'^{0}$.  

Also in (A5a), $b^{\dagger}(k)b(k)+b(k)b^{\dagger}(k)$ has been replaced by $2b^{\dagger}(k)b(k)$.   When $\nu\neq 0$, the infinite contribution of the commutator vanishes since $\sim \int d{\bf k}{\bf k}$ vanishes. When $\nu= 0$, we disregard the infinite vacuum energy as is usual.  We also note  that (A5a)'s curly bracketed term simply becomes $k^{0}k^{\nu}$, since the delta and step functions imply that $k=k'$.

	In going from (A5a) to (A5b), we have performed the integrals over $k'$ and  $m^{2}$.  The integral over $k'^{0}$ of $\delta(k^{2}-k'^{2})\rightarrow\delta[(k^{0}-k'^{0})(k^{0}+k'^{0})]$ produces the factor $1/(2|k^{0}|)$.  The integral over $m^{2}$ provides the step function limit on the range of $k^{0}$ in (A5b). 

	In going from (A5b) to (A5c), we have used $\sum_{s}sk^{0}\Theta(sk^{0})=|k^{0}|$. 
	
	It now remains to express $T_{w}^{\mu\nu}(x)$ in terms of the random fields $W(x)$ and $\Pi(x)$,  instead of the fields $\phi_{m,r,s}(x)$.  To that end we apply (\ref 9) to (\ref{A1}), obtaining: 
\begin{equation}\label{A6}
\phi_{m,r,s}(x)=\frac{1}{(2\pi)^{7/2}}\int dx_{1} \int dk \delta (k^{2}+rm^{2})\Theta (sk^{0})
[\frac{1}{\sqrt{\lambda}}W(x_{1})\cos k\cdot(x-x_{1})-2\sqrt{\lambda}\Pi(x_{1})\sin k\cdot(x-x_{1})],
\end{equation}
and substitute this into (\ref{A3}):
\begin{eqnarray}\label{A7}
T_{w}^{\mu\nu}(x)&=&\frac{1}{(2\pi)^{7}}\sum_{rs}\int_{0}^{\infty}dm^{2}s\int dx_{1}dx_{2} \int dk_{1} dk_{2}
\delta (k_{1}^{2}+rm^{2})\delta (k_{2}^{2}+rm^{2})\Theta (sk_{1}^{0})\Theta (sk_{2}^{0})\nonumber\\
&&\Bigg\{\Big\{ [k_{1}^{\mu}k_{2}^{\nu}-(1/2)\eta^{\mu\nu}(k_{1}\cdot k_{2}+k_{2}^{2})]\Big[\frac{1}{\sqrt{\lambda}}W(x_{1})\sin k_{1}\cdot(x-x_{1})+2\sqrt{\lambda}\Pi(x_{1})\cos k_{1}\cdot(x-x_{1})\Big]\nonumber\\
&&\qquad\cdot\Big[\frac{1}{\sqrt{\lambda}}W(x_{2})\sin k_{2}\cdot(x-x_{2})+2\sqrt{\lambda}\Pi(x_{2})\cos k_{2}\cdot(x-x_{2})\Big]\Bigg]\nonumber\\
&&+(1/2)\eta^{\mu\nu}k_{2}^{2}
\Big[\frac{1}{\sqrt{\lambda}}W(x_{1})\cos k_{1}\cdot(x-x_{1})-2\sqrt{\lambda}\Pi(x_{1})\sin k_{1}\cdot(x-x_{1})\Big]\nonumber\\
&&\qquad\cdot\Big[\frac{1}{\sqrt{\lambda}}W(x_{2})\cos k_{2}\cdot(x-x_{2})-2\sqrt{\lambda}\Pi(x_{2})\sin k_{2}\cdot(x-x_{2})\Big]\Bigg\}.
\end{eqnarray}

When $m^{2}$ is integrated over and $r$ is summed over, the range of integration of $k_{1}, k_{2}$ becomes unrestricted. 
When $s$ and the variables of integration $k_{1}, k_{2}$ are changed in sign in (\ref{A7}), the result is 
(\ref{A7}) with the overall sign changed and with the sign of the sine's reversed.  When half of this is added to half  of (\ref{A7}), the coefficients of $W(x_{1})W(x_{2})$ and $ \Pi(x_{1})\Pi(x_{2})$ vanish. Moreover, the result is unaltered if we add, to half of it, half the same expression with the variables of integration  $k_{1}, k_{2}$ exchanged and $x_{1}, x_{2}$ exchanged, obtaining
\begin{eqnarray}\label{A8}
T_{w}^{\mu\nu}(x)&=&\frac{1}{2(2\pi)^{7}}\sum_{s}s\int dx_{1}dx_{2} \int dk_{1} dk_{2}
\delta (k_{1}^{2}-k_{2}^{2})\Theta (sk_{1}^{0})\Theta (sk_{2}^{0})\nonumber\\
&&\Big\{ [k_{1}^{\mu}k_{2}^{\nu}+k_{1}^{\nu}k_{2}^{\mu}-(1/2)\eta^{\mu\nu}(k_{1}+ k_{2})^{2}]
\sin [k_{1}\cdot(x-x_{1})+k_{2}\cdot(x-x_{2})][W(x_{1})\Pi(x_{2})+\Pi(x_{1})W(x_{2})]\nonumber\\
&&\negthinspace\negthinspace\negthinspace\negthinspace\negthinspace\negthinspace\negthinspace\negthinspace
+[k_{1}^{\mu}k_{2}^{\nu}+k_{1}^{\nu}k_{2}^{\mu}+(1/2)\eta^{\mu\nu}(k_{1}-k_{2})^{2}]
\sin [k_{1}\cdot(x-x_{1})-k_{2}\cdot(x-x_{2})][W(x_{1})\Pi(x_{2})-\Pi(x_{1})W(x_{2})]\Big\}.
\end{eqnarray}

Now, the first term within the curly brackets in (\ref{A8}) does not contribute to $P_{w}^{\nu}$.  For it,  
$\int d{\bf x}T_{w}^{0\nu}(x)$ provides an overall factor of $\delta({\bf k}_{1}+{\bf k}_{2})$, so ${\bf k}_{1}=-{\bf k}_{2}$.  This, together with the other delta function and step functions gives $k_{1}^{0}= k_{2}^{0}$.  Thus, for  $P_{w}^{0}$, the first factor in the first  term within the curly brackets becomes
\[ k_{1}^{0}k_{2}^{0}+k_{1}^{0}k_{2}^{0}-(1/2)\eta^{00}(k_{1}+ k_{2})^{2}=2(k_{1}^{0})^{2}-(1/2)(2k_{1}^{0})^{2}=0
\]  
while for $P_{w}^{i}$, this factor becomes
\[k_{1}^{0}k_{2}^{i}+k_{1}^{i}k_{2}^{0}=k_{1}^{0}(k_{1}^{i}-k_{1}^{i})=0.
\] 
Therefore, for simplicity we may omit this term, without affecting any of the needed properties of $T_{w}^{\mu\nu}(x)$
Moreover, one readily sees that 
\[[W(x_{1})\Pi(x_{2})+\Pi(x_{1})W(x_{2})]\sim [W_{+}(x_{1})W_{+}(x_{2})-W_{-}(x_{1})W_{-}(x_{2})]
\]
where $W_{+}(x)$ is the positive frequency (annihilation) part of $W(x)$ and $W_{-}(x)$ is its negative frequency (creation) part.  Thus, by omitting this term, we eliminate the complication of  $T_{w}^{\mu\nu}(x)$ creating or annihilating pairs of  ``W-particles."  This is not the case with the second term we keep, since 
\[[W(x_{1})\Pi(x_{2})-\Pi(x_{1})W(x_{2})]\sim [W_{+}(x_{1})W_{-}(x_{2})-W_{-}(x_{1})W_{+}(x_{2})].
\]
which maintains the number of ``W-particles." 

Therefore, our result is:  
\begin{eqnarray}\label{A9}
T_{w}^{\mu\nu}(x)&=&\frac{1}{2(2\pi)^{7}}\int dx_{1}dx_{2} \int dk_{1} dk_{2}
\delta (k_{1}^{2}-k_{2}^{2})\sum_{s}s\Theta (sk_{1}^{0})\Theta (sk_{2}^{0})\sin [k_{1}\cdot(x-x_{1})-k_{2}\cdot(x-x_{2})]\nonumber\\
&&[k_{1}^{\mu}k_{2}^{\nu}+k_{1}^{\nu}k_{2}^{\mu}+(1/2)\eta^{\mu\nu}(k_{1}-k_{2})^{2}]
[W(x_{1})\Pi(x_{2})-\Pi(x_{1})W(x_{2})].
\end{eqnarray}
which is further discussed in section II.

\section{Form Factor}

In this appendix we calculate the form factor which appears in Eq.(11b), 
which expresses the non-local nature of the stress tensor: 
\begin{equation}\label{B1}
G(x-x_{1}, x-x_{2})\equiv\frac{1}{(2\pi)^{7}}\int dk_{1} dk_{2}
\delta (k_{1}^{2}-k_{2}^{2})\sum_{s}s\Theta (sk_{1}^{0})\Theta (sk_{2}^{0})\sin [k_{1}\cdot(x-x_{1})-k_{2}\cdot(x-x_{2})].
\end{equation}

	To evaluate $G$, we look at 	
\begin{equation}\label{B2}
g(s_{1}, s_{2})\equiv\frac{1}{i(2\pi)^{7}}\int dk_{1} dk_{2}
\delta (k_{1}^{2}-k_{2}^{2})\epsilon(k_{1}^{0})e^{i[k_{1}\cdot s_{1}-k_{2}\cdot s_{2}]},
\end{equation}	
where $s_{i}\equiv (x-x_{i})$, and note that 
\[
g(s_{1}, s_{2})=G(s_{1}, s_{2})+\frac{2}{(2\pi)^{7}}\int d{\bf k}_{1} d{\bf k}_{2}
\delta (k_{1}^{2}-k_{2}^{2})\int_{0}^{\infty} dk_{1}^{0}\int_{0}^{\infty} dk_{2}^{0}\sin[k_{1}\cdot s_{1}+k_{2}\cdot s_{2}]
\]
so
\begin{equation}\label{B3}
G(s_{1}, s_{2})=\frac{1}{2}[g(s_{1}, s_{2})-g(s_{2}, s_{1})].
\end{equation}

To evaluate $g$, we note that,  in (\ref{B2}) , it is 
\[
\epsilon(k_{1}^{0})=\frac{1}{2i\pi }\int d\Omega \frac {\Omega}{\Omega^{2}+\Delta^{2}}e^{i\Omega k_{1}^{0}}
\]
($\Delta$ is an infinitesimal) which keeps the expression from being relativistically invariant and easy to evaluate.  Therefore, we take $\epsilon(k_{1}^{0})$ out of the integral:  with $T_{i}\equiv s_{i}^{0}$, we write (\ref{B2}) as
\begin{eqnarray}\label{B4} 
g(s_{1}, s_{2})&=&\frac{1}{2i\pi }\int d\Omega \frac {\Omega}{\Omega^{2}+\Delta^{2}}e^{-\Omega \partial/\partial T_{1}}
\frac{1}{i(2\pi)^{7}}\int dk_{1} dk_{2}
\frac{1}{2\pi}\int d\omega e^{i\omega [k_{1}^{2}-k_{2}^{2}]}e^{i[k_{1}\cdot s_{1}-k_{2}\cdot s_{2}]}\nonumber\\
&=& \frac{1}{2i\pi }\int d\Omega \frac {\Omega}{\Omega^{2}+\Delta^{2}}e^{-\Omega \partial/\partial T_{1}}
\frac{\pi^{4}}{i(2\pi)^{8}}\int d\omega \frac{1}{\omega^{4}}e^{-i(s_{1}^{2}-s_{2}^{2})/4\omega}\nonumber\\
&=&-\frac{1}{8\pi^{5}}\int d\Omega {\cal P}\frac{1}{\Omega} \int d\omega '\omega '^{2}e^{-i\omega '
[\sigma+T_{1}^{2}-(T_{1}-\Omega)^{2}]}=\frac{1}{4\pi^{4}}\frac{\partial^{2}}{\partial\sigma^{2}}\int d\Omega {\cal P}\frac{1}{\Omega} \delta[\sigma+T_{1}^{2}-(T_{1}-\Omega)^{2}]\nonumber\\
&=&-\frac{1}{4\pi^{4}}\frac{\partial^{2}}{\partial\sigma^{2}}\Bigg\{
\frac{T_{1}\Theta (\sigma+T_{1}^{2})}{\sqrt{\sigma+T_{1}^{2}}}{\cal P}\frac{1}{\sigma}\Bigg\}
\end{eqnarray}
where we have written $\sigma\equiv s_{1}^{2}-s_{2}^{2}$, and noted that $\Omega/(\Omega^{2}+\Delta^{2})={\cal P}1/\Omega$ (the principal value).  

	Thus, we conclude, from (\ref{B3}), (\ref{B4}),
\begin{equation}\label{B5}
G(s_{1}, s_{2})=-\frac{2}{(2\pi)^{4}}\frac{\partial^{2}}{\partial\sigma^{2}}\Bigg\{\Bigg[
\frac{T_{1}\Theta (\sigma+T_{1}^{2})}{\sqrt{\sigma+T_{1}^{2}}}+\frac{T_{2}\Theta (-\sigma+T_{2}^{2})}{\sqrt{-\sigma+T_{2}^{2}}}\Bigg]{\cal P}\frac{1}{\sigma}\Bigg\}.
\end{equation}

\end{document}